\documentclass [12pt]{iopart}
\usepackage{graphicx}

\usepackage{iopams}  

\begin{document}

\title{The sonic analogue of black hole radiation}
\author{S. Giovanazzi}
\address{Institut f\"{u}r Theoretische Physik III und 5 Physikalisches Institut, Universit\"{a}t Stuttgart, 70550 Stuttgart, Germany}
\ead{stevbolz@yahoo.it} 

\begin{abstract}
A microscopic description of Hawking radiation in sonic black holes has been recently presented \cite{Giovanazzi05a}. 
This exactly solvable model is formulated in terms of one-dimensional scattering of a Fermi gas.
In this paper, the model is extended to account possible finite size effects of a realistic geometry. 
The flow of particles is maintained by a piston (i.e. an impenetrable barrier) moving slowly towards the sonic horizon. 
Using existing technologies the Hawking temperature can be of order of a few microkelvin in a realistic experiment.
\end{abstract}
\noindent{\it Keywords}: Hawking radiation, sonic black holes, quantum tunnelling, ultra-cold gases
\pacs{03.75.Ss, 04.70.Dy}
\submitto{\JPB}
\maketitle

\section{Introduction} 
Analogue models for black holes \cite{BookABH} are recently subject of a renewed interest \cite{Giovanazzi05a,BookABH,Visser98,Garay00,Garay01,Leonhardt02,Barcelo03,Fischer03,Leonhardt03a,Leonhardt03b,Giovanazzi04}
since the realization of atomic Bose-Einstein condensate \cite{Cornell02,Ketterle02} and Fermi degenerate gases \cite{Demarco99,Schreck01} at temperature of a few tens of nanokelvin. 
These achievements together with the realization of one-dimensional systems \cite{Cassettari99,Folman00,Ott01,Hansch01,Goerlitz01,Greiner01} make the experimental realization in the laboratory of an analogue of Hawking radiation \cite{Unruh81} possible. This possibility has been made clear since the understanding of the relationship of sonic Hawking radiation with ``the tunnelling through the top of the hill'' \cite{Giovanazzi05a}.

Analogue models for black holes have been originally considered in the hope that they can shed some light into the problem of quantization of gravity and on the related problem of information loss \cite{Hawking05}. In 1974, Hawking \cite{Hawking74} proved that black holes can emit particles with a temperature proportional to its surface gravity. The predicted spectrum of the radiation is purely thermal and incoherent and this implies a loss of information after the black hole has evaporated away and disappeared completely. 
In 1981, Unruh \cite{Unruh81} suggested that an analogue of Hawking radiation can be produced in a quantum fluid turning supersonic. Very recently, a microscopic theory for this sonic analogue of Hawking radiation has been presented \cite{Giovanazzi05a}. This theory, although unable to describe the process of evaporation of the black hole analogue, is based on elementary quantum mechanics and thus, it is unitary and information preserving. 

In this paper, I review the microscopic model for sonic Hawking radiation (Section 2) and extend the model to a finite size geometry where the source of the particles is replaced by a moving piston, i.e. an impenetrable barrier (Section 3). This problem bears analogies with the ``Fermi accelerator problem'' \cite{Fermi49}. The possibility of an experimental realization of the model is discussed in Section 4.
Session 5 concludes.

\section{Microscopic description} 
A one-dimensional Fermi-degenerate non-interacting gas that scatters against a very smooth potential barrier provides a clear and simple quantum mechanical microscopic description of the sonic analogue of Hawking radiation.
Sonic Hawking radiation is formed due to quantum tunnelling through the top of the potential barrier. Excitations, in a semiclassical sense are created at the barrier and are then radiated with a momentum distribution characterized by a temperature in exact agreement with Hawking-Unruh's formula \cite{Hawking74,Unruh81,Visser98}. Although the evolution of the fluid's flow is unitary, when the momentum distribution is measured locally, for instance through the measurement of the dynamic structure factor, it is not distinguishable from a thermal (and incoherent) distribution. 

\subsection{One-dimensional systems}
Consider a Fermi-degenerate gas of particles moving in a sufficiently narrow and smooth constriction such that their motion can be considered  one-dimensional.
Their stationary single-particle wave-functions 
can be written in the adiabatic approximation \cite{Glazman88} as $\Psi(x,y,z)=\psi(x)\xi_{0;x}(y,z)$ where $\xi_{0;x}(y,z)$ is the transverse ground-state wave-function with eigenenergy $\epsilon_{0}(x)$. $\psi(x)$ satisfies a  one-dimensional Schr\"{o}dinger equation 
$[-(\hbar^{2}/2m)\partial_{x}^{2}+ V_{\rm ext}(x)]\psi(x)=
\epsilon\psi(x)$ where $V_{\rm ext}$ is an effective potential that includes also the zero point energy $\epsilon_{0}(x)$, i.e. 
\begin{equation}
V_{\rm ext}(x)= \langle\xi_{0;x}| -(\hbar^{2}/2m)(\partial_{y}^{2}+\partial_{z}^{2})+ V_{\rm ext}(x,y,z)|\xi_{0;x} \rangle
\label{vext}
\end{equation}
where $V_{\rm ext}(x,y,z)$ is the three dimensional external potential.
I assume $V_{\rm ext}(x)$ very smooth and essentially non-zero only near its maximum $V_{\rm max}$ located at $x=0$.

\subsection{Unitary description}
The transonic flow has an unitary quantum mechanical description.
Consider a flow of  non-interacting particles coming from the left and colliding against the potential $V_{\rm ext}$.
Their particle's wave-functions are given by scattering wave functions asymptotically defined by  $\psi_{\epsilon} \propto  [ \rme^{i k x} + r(\epsilon) \rme^{- i k x} ] \Theta(-x)+ t(\epsilon) \rme^{i k x}  \Theta(x) $, where $\epsilon=\hbar^{2}k^{2}/2m$ is the energy eigenvalue and $r(\epsilon)$ and $t(\epsilon)$  are the reflection and transmission amplitudes, respectively.
The evolution of the particle flow is unitary and the many-body wave function of the flow of particles is represented in second quantization by
\begin{equation}
| f \rangle = \prod_{0 \le \epsilon < \epsilon_{\rm max}} a^{\dag}_{\epsilon}| - \rangle
\label{flow}
\end{equation}
where $a^{\dag}_{\epsilon}$ is the creation operator of $\psi_{\epsilon}$'s and $| - \rangle$ is the vaccum of particles. In this state all $\psi_{\epsilon}$'s are occupied up to an energy $\epsilon_{\rm max}$, which is assumed higher than $V_{\rm max}$.
The non-interacting case is an ideal example of a fluid turning supersonic (see below) where the microscopic wave function of the flow (\ref{flow}) is given from the beginning.

\subsection{Hydrodynamic description}
Since the barrier potential is smooth the particle propagation is semi classical and in first approximation $|t(\epsilon)|^{2} \approx \Theta(\epsilon-V_{\rm max})$.
Therefore the fluid maintains in a first approximation the sharp Fermi-degenerate character of the local momentum (or velocity) distribution, which represents the ``Vacuum'' of the black hole sonic analogue.
This allows us to describe the fluid dynamic with the standard hydrodynamic equations.
The velocities of the particles in the fluid are uniformly distributed within an interval $(v_{L},v_{R})$. Figure 1 shows an example of velocity distribution. The right Fermi velocity 
\begin{equation}
v_{R}=\sqrt{v_{\rm max}^{2} - 2 V_{\rm ext}(x)/m}
\end{equation}
 is the velocity of a particle moving from the left to the right with initial velocity higher than the classical threshold velocity to go over the barrier 
\begin{equation}
 v_{\rm esc}=\sqrt{2 V_{\rm max}/m}
\end{equation}
The left Fermi velocity $v_{L}$ is the velocity of a particle starting from the top of the barrier (located at $x=0$), i.e. 
\begin{equation}
v_{L}= {\rm sgn}(x) \; \sqrt{v_{\rm esc}^{2} - 2 V_{\rm ext}(x)/m}.
\end{equation}
The ${\rm sgn}(x)$ function accounts for the two possible different signs of the velocity that the particle can have falling either to the right or left of the barrier.
\begin{figure}[htbp]
\begin{center}
\includegraphics[width=100mm]{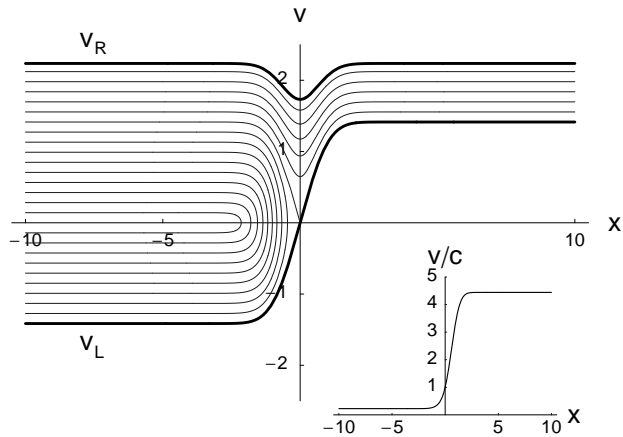}
\caption{Semi classical velocity distribution as function of the coordinate $x$. Velocities are uniformly distributed in the area limited by 
$v_{R}$ (top solid line) and $v_{L}$ (bottom solid line). Thinner solid lines correspond to different classical trajectories of the particles of the fluid.
Here $\epsilon_{\rm max} = 2.5$ and $V_{\rm ext}  = \exp(-x^{2})$ in dimensionless units. The inset shows the Mach number $v/c$ as function of position $x$.}
\label{fig2}
\end{center}
\end{figure}

The flow velocity $v=(v_{L}+v_{R})/2$ is the velocity of a local reference frame from where the fluid appears in equilibrium. Introducing the Fermi velocity $v_{F}=(-v_{{\rm L}} + v_{{\rm R}})/2$ as the Fermi velocity in the co-moving frame,  the density is given by $n=m v_{F} / \pi\hbar$. 
With these definitions the current $\Omega = n v=m v v_{F} / \pi\hbar$ is conserved. [The current conservation can be written in a form that is reminescent of the quantization of the conductance in  one-dimensional channels: $\Omega = (\mu_R - \mu_L)/ h_{{\rm P}}$ where $h_{{\rm P}}$ is the Planck constant, $\mu_{L}= m v_{\rm esc}^{2}/2$ and $\mu_{R}= m v_{\rm max}^{2}/2 $.] 
The Bernoulli equation $v^{2}/2 +  h + V_{\rm ext}/m={\rm const}$ where $h(n)=\pi^2\hbar^2 n^2/2m^{2}$ is
the enthalpy (the Fermi energy divided by the particle mass) is satisfied with ${\rm const}=v^{2}_{\rm esc}/2+v^{2}_{\rm max}/2$. 
The sound velocity ($c^2=n \,dh(n) / dn$) is consistently given by the Fermi velocity $c = v_{F}$. It is easy to verify that the above defined particles flow is a transonic flow, namely subsonic on the left of the barrier and supersonic on the right. The inset of Figure 1 shows the behaviour of the Mach number $v/c$ in the  one-dimensional channel.

Although, here I showed that the stationary hydrodynamic equations are satisfied for the gas flow, it is important to state that also the long-wave perturbations of the stationary solution, i.e. sound waves satisfy the time-dependent hydrodynamic equations in the limit of very small amplitudes. This will be reported in details elsewhere. This condition is important in order to apply Unruh's theory for sonic black holes \cite{Unruh81}.

\subsection{Quantum effects: the temperature of sonic analogue of Hawking radiation}
Even if the potential barrier is very smooth, its finite thickness near its top smears the semi classical Fermi distributions at $v_{{\rm L}}(x)$. This is due to those particles colliding against the barrier with energies comparable with $V_{\rm max}$.
The reflection probability is given by (see also \cite{Hanggi90}) 
\begin{equation}
|r(\epsilon)|^{2}=\frac{1}{1+\exp(2\pi(\epsilon-V_{\rm max})/\hbar\omega_{x})}
\label{penetration}
\end{equation}
where $\omega_{x}$ is the frequency of the inverted harmonic oscillator obtained expanding the barrier potential $V_{\rm ext} = V_{\rm max} - \frac 12 m \omega_{x}^2 x^2 + o(x^3)$ around its maximum. 
Hawking's formula for a sonic black hole is given by \cite{Unruh81,Visser98}
\begin{equation}
T_H=\left( \frac{\hbar}{2\pi k_B} \right) \frac{d(v-c)}{dx} \,,
\label{unruh}
\end{equation}
where the surface gravity $\kappa$ in Hawking temperature $\kappa\hbar / 2\pi k_B$ is replaced by 
$d(v-c)/dx$. For the transonic flow of fermions defined above $d(v-c)/dx=d(v_{L})/dx=\omega_{x}$ and Hawking temperature (\ref{unruh}) becomes
\begin{eqnarray}
T_{{\rm H}}= \frac{\hbar \omega_{x}}{2\pi k_{{\rm B}}}  \;.
\label{bare1}
\end{eqnarray}
Therefore the particles reflected from the barrier are distributed according to the reflection probability (\ref{penetration}), which is identical to a thermal Fermi distribution 
\begin{equation}
n_{T,\mu}(\epsilon)=\frac{1}{1+\exp((\epsilon-\mu)/k_{{\rm B}}\,T)}
\label{thermaldistribution}
\end{equation}
with a chemical potential $\mu=V_{\rm max}$ and a temperature in exact agreement with Hawking temperature (\ref{bare1}).

When quantum tunneling is negligible the quantum state observed over a large segment $L$ far away from the barrier would appear like a zero temperature Fermi-degenerate gas, which can be viewed as the semiclassical vacuum of an effective low-energy description. The role of the tunneling through the top of the barrier is to introduce a smoothed Fermi left point in $L$ and thus create excitations on the semiclassical vacuum.

We note that this interpretation of sonic Hawking radiation as related to a tunnel effect is consistent with the recent derivation of Hawking radiation as a tunnelling process of particles in a dynamical geometry due to Parikh and Wilczek \cite{Wilczek00}. 

Moreover as quantum tunnelling is very sensitive to the details of the potential barrier, the momentum distribution, related to the reflection coefficient of the barrier, can be different from the inverted parabola formula (\ref{penetration}). 
This suggest that some information on the internal geometry of the black hole is carried outside by the radiation.

\subsection{Measuring the sonic Hawking radiation}

Sonic Hawking radiation can be detected through the measurement of the dynamic structure factor $S(q,\omega)$ \cite{Pines} in an extended segment of the subsonic or supersonic region  \cite{Giovanazzi05a}.
The $S(q,\omega)$ of the flow is given \cite{Giovanazzi05a} in the subsonic region by 
\begin{eqnarray}
S(q,\omega)=S^{{\rm eq}}_{0,\mu_{R}}(q,\omega)\,\Theta\left(\frac{\omega}{q}\right)+S^{{\rm eq}}_{T_{{\rm H}},\mu_{L}}(q,\omega)\,\Theta\left(-\frac{\omega}{q}\right)
\end{eqnarray}
and in the supersonic region by 
\begin{eqnarray}
S(q,\omega)=\left[S^{{\rm eq}}_{0,\mu_{R}}(q,\omega)+S^{{\rm eq}}_{T_{{\rm H}},\mu_{L}}(q,-\omega)\right]\Theta\left(\frac{\omega}{q}\right)
\end{eqnarray}
 with $\mu_{L}=V_{\rm max}$ and $\mu_{R}=\epsilon_{\rm max}$. 
$S^{{\rm eq}}_{T,\mu}(q,\omega)$ is the dynamic structure factor of an equilibrium  one-dimensional Fermi gas at temperature $T$ and chemical potential $\mu$ given by $S^{{\rm eq}}_{T,\mu}(q,\omega)= \frac{L_{{\rm eff}}\, m}{4\pi\hbar^{2}q }\, e^{\hbar\omega/k_{{\rm B}}T}/\{{\rm Cosh}(\hbar\omega/k_{{\rm B}}T)+
{\rm Cosh}[(\frac{m\omega^{2}}{2q^{2}}+\frac{\hbar^{2} q^{2}}{8 m} -\mu)/k_{{\rm B}}T]\}$
where $L_{{\rm eff}}$ is an effective length of the probe-system interaction region \cite{Giovanazzi05a}. 
Its $T=0$ limit is given by 
$S^{eq}_{0,\mu}(q,\omega)=\frac{L_{{\rm eff}}\, m}{2\pi\hbar^{2}q }\,\Theta\left(-\omega+\frac{\hbar^{2}q^{2}}{2m}+\frac{\hbar^{2}q k_{F}^{o}}{m}\right)\Theta\left(\omega+\frac{\hbar^{2}q^{2}}{2m}-\frac{\hbar^{2}q k_{F}^{o}}{m}\right)$ with $k_{F}^{o}=\sqrt{2 m \mu}/\hbar$. 

Figure 2 shows some examples of $S(q,\omega)$.
$S(q,\omega)$ has a typical peaks structure at low momentum, which contrasts the non-interacting two and three dimensional Fermi gas \cite{Pines}. The peaks are related to the dispersion relation $\omega(q)=-v_{{\rm L}} \,|q|\,\Theta(-q)+v_{{\rm R}}  \,|q|\,\Theta(q)$ valid for small $q$. 
The characteristic wave-vectors of Hawking radiation is $q_{c}= k_{{\rm B}}T_{{\rm H}}/\hbar c\sim1/2\pi^{2}l_{x}^{2}n$ where $l_{x}=\sqrt{m/\hbar\omega_{x}}$ and $n$ is the density. Indeed
at very low frequencies $\hbar\omega \ll k_{{\rm B}}T_{{\rm H}}$ and momentum $q\ll q_{c}$ the $S(q,\omega)$ is dominated by the  Hawking radiation. 
The peak values at $|\omega|=|v_{{\rm L}}q| $ tends to $1/4$ of the corresponding zero-temperature value $L_{{\rm eff}}\, m\,/\,2\pi\hbar^{3}q $ but with a half-width $\Delta\omega=1.76\, q k_{{\rm B}}T_{{\rm H}} / m v_{{\rm L}}$, which is much larger than the corresponding half-width $\Delta\omega=\hbar q^{2} / 2 m $ of the zero-temperature contribution. 
The measure of the width of these peaks provides a measure of the Hawking temperature of the sonic black hole.

\begin{figure}[htbp]
\begin{center}
\includegraphics[width=100mm]{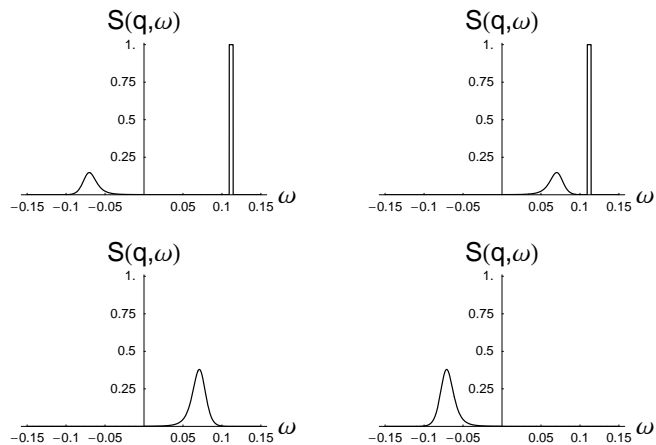}
\caption{Sonic Hawking radiation can be detected through the measurement of $S(q,\omega)$. 
The figure shows the $\omega$-dependence of $S(q,\omega)$ in the subsonic (left) and supersonic (right) regions and for a positive (top) and a negative (bottom) value of $q$. Here $k_{\rm B}T_{\rm H} = 0.15\, V_{\rm max}$, $\epsilon_{\rm max} = 2.5 \, V_{\rm max}$ and $|q|=0.05\,m v_{\rm esc}/\hbar$ is chosen in the region of small $q$ where thermal-like transitions (broadened peaks) dominate over the zero-temperature contributions (step-like peaks). $S(q,\omega)$ is in units of $L_{\rm eff}\, m\,/\,2\pi\hbar^{3}q $ and $\omega$ is in units of $V_{\rm max}/\hbar$. }
\label{figsqo}
\end{center}
\end{figure}

This calculation shows also that a local measurement of $S(q,\omega)$  through a scattering experiment in the subsonic or supersonic regions does not probe the underline global coherence of the flow.
The $S(q,\omega)$ in the subsonic region is equal to that of a clean one-dimensional channel connecting two reservoirs: one on the left at zero temperature and one on the right at the Hawking temperature $T_{{\rm H}}$ given by (\ref{bare1}).
This represents a one-dimensional analogue of Hawking radiation from a black hole into the surrounding zero-temperature space.

To distinguish between the pure quantum state of this fluid, which is represented by the many-particle wave function (\ref{flow}) and a mixed thermal state using only measurements made in the subsonic or in the supersonic region is probable difficult. 
This  would require the measurement of correlations such as those between opposite sides of the barrier. When a particle is detected in the subsonic region at some negative momentum $k$, i.e. reflected by the barrier, we know that it will not be detected in the supersonic region at momentum $-k$.

\section{The piston geometry}

In the model introduced in Section 2 the quantum mechanical description of the many-body system is given in terms of single-particle  scattering wave-functions. This is a very ideal situation where we theoretically identified the predicted Hawking radiation. In a practical realization of the model, one can perhaps imagine to join the one-dimensional channel with a large reservoir on one side the channel to ensure a continuum source for degenerate Fermi particles and with an empty area on the other side where the atoms can escape. One needs to assume that in the reservoir the interaction of the particles is sufficiently effective to ensure that the quantum mechanical statistical equilibrium takes place quickly enough. Therefore, the {\it complete} quantum mechanical description of the entire setup is not possible in simple terms and the local quantum mechanical description in the one-dimensional channel relies on assumptions on the reservoir.

In the following we consider a finite size system as a model for a sonic black hole that can be fully described quantum mechanically and {\it no assumption of equilibration is necessary}. The reservoir is replaced by a time-dependent impenetrable potential barrier (the ``piston'') that with his constant motion pushes the gas particles from the subsonic region over a smooth barrier in the supersonic region. The transonic flow will take place at the top of the smooth barrier.

\subsection{Hydrodynamic description}

The piston, i.e. an impenetrable and sharp potential barrier of coordinate $x_{\rm piston}(t)<0$ moves at constant velocity $v_{\rm piston}>0$ in a region of zero potential towards the barrier located at $x = 0$. The Fermi-degenerate gas is initially trapped between the piston and barrier, i.e. in the subsonic region. The piston compresses the gas, creating thus the conditions of pressure to start the transonic flow. The gas will later flow stationary for a long time and its velocity $v$ in the subsonic region will be equal to the piston velocity, i.e. $v = v_{\rm piston}$.

In the initial part of the gas dynamic, i.e. before that the atoms start to escape to the right of the barrier, the gas particles have an uniform semiclassical distribution of velocity in the subsonic region given approximatively by
\begin{eqnarray}\label{fxvt}
f(x,v,t)\,\rmd x \rmd v = \frac m{h_{\rm P}} \,\Theta[-v_{\rm L}(x,t)-v]\, \Theta[v_{\rm R}(x,t)-v]\, \rmd x \rmd v
\end{eqnarray}
where $h_{\rm P}$ is the Planck constant.
Initially the local left and right Fermi velocities $v_{\rm L}(x,t)$ and $v_{\rm R}(x,t)$ have modulus less than the escape velocity  $v_{\rm esc}$ and are given by
\begin{eqnarray}\label{vli}
v_{\rm L}(x,t)&=&-v_{\rm F}(t)-v_{\rm piston} \frac{x}{|x_{\rm piston}(t)|}\\\label{vri}
v_{\rm R}(x,t)&=&v_{\rm F}(t)-v_{\rm piston} \frac{x}{|x_{\rm piston}(t)|}
\end{eqnarray}
where
\begin{equation}\label{pi}
x_{\rm piston}(t) = x_{\rm piston}(0)+ v_{\rm piston}\, t
\end{equation}
The Fermi velocity $v_{\rm F}(t)$ is initially fixed by the conservation of the number of fermions $N_{\rm F}$ in the left side region
and is given approximatively by
\begin{equation}\label{vfi}
v_{\rm F}(t) \approx {N_{\rm F} {\pi \hbar} \over m |x_{\rm piston}(t)|}
\label{ndopoini}
\end{equation}
We assume for simplicity that the barrier size is much smaller than $|x_{\rm piston}(t)|$. 
The finite size of the barrier results in a modification of the transient dynamics, which discussion is beyond the purpose of this paper, and does not concern the quantum effect occurring at the sonic event horizon. 
 
Every time that one atom bounces against the moving piston its absolute value of the velocity increases by $2 v_{\rm piston}$. 
At a time $t_{trans}$ some  of the particles near the barrier reach the escape velocity $v_{\rm esc}$ 
and they start to go over the barrier in the supersonic region. After a transient time of about $2 |x_{\rm piston}(t_{trans})|/v_{\rm esc}$ the gas will enter a stationary regime characterized by a semiclassical distribution of velocity with a form still given by equation (\ref{fxvt}) but with constant left and right Fermi velocities $v_{\rm L}$ and $v_{\rm R}$ given in the subsonic region by
\begin{eqnarray}
v_{\rm L}&=&-v_{\rm esc}\\
v_{\rm R}&=&v_{\rm esc} + 2 v_{\rm piston}
\end{eqnarray}
and in the supersonic region by
\begin{eqnarray}
v_{\rm L}&=&v_{\rm esc}\\
v_{\rm R}&=&v_{\rm esc} + 2 v_{\rm piston}
\end{eqnarray}
Figure 3 sketches the semiclassical Fermi distribution during the stationary phase  of the gas dynamics.
Note that the particles that collide with the piston at velocity $v_{\rm L}$ are reflected back at velocity $v_{\rm R}$. Thus the moving piston
connects $v_{\rm L}$ with $v_{\rm R}$.

\begin{figure}[htbp]
\begin{center}
\includegraphics[width=120mm]{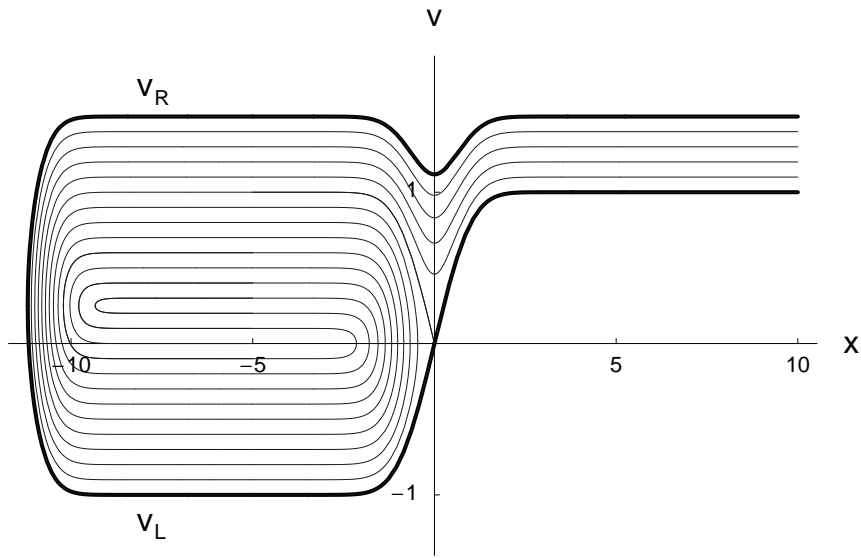}
\caption{Semi-classical velocity distribution as function of the coordinate $x$ for a certain time chosen in the stationary regime. Velocities are uniformly distributed in the area limited by the solid line. Thinner solid lines correspond to different classical trajectories of the particles of the fluid. 
Here $V_{\rm ext}  = (1/2) \exp(-x^{2})$, $v_{\rm esc} = 1$ and 
$v_{\rm piston} = 0.25$  in dimensionless units.
Note that the piston ($x \approx -12$) moving to the right recombines the left Fermi velocity $v_{L}$ (bottom part of the solid line) with the right Fermi velocity $v_{R}$ (top part of the solid line) in the subsonic region. [For the purpose of rendering the particles trajectories in this ``instantaneous'' picture it is necessary to assume that the collisions of the particles with the piston take place instantaneously.] }
\label{fig3}
\end{center}
\end{figure}

The gas flow will eventually terminate when the piston reaches the barrier after the time interval $|x_{\rm piston}(t_{trans})|/v_{\rm piston}$. 
The above stationary transonic flow can also form starting with different initial conditions and these will affect mainly the transient dynamics. For instance, the transonic regime can start also by lowering the barrier high.

\subsection{Quantum effects of the sonic event horizon}

The next step is to consider the quantum description of the gas dynamics and the consequences of the tunneling at the top of the barrier.
The quantum state of the ${\rm N}_{\rm F}$ particles gas is given by 
\begin{equation}
| f \rangle = \prod_{1 \le {\rm n} \le {\rm N}_{\rm F}} a^{\dag}_{\rm n}| - \rangle
\label{flow2}
\end{equation}
where $| - \rangle$ is the vaccum of particles and $a^{\dag}_{\rm n}$ creates one fermion in the single-particle state $\psi_{\rm n} (x,t)$, which is  solution of the time-dependent Schr{\"o}dinger equation. 
In the initial stage of the dynamics $\psi_{\rm n} (x,t)$ are quasi eigenstates  (resonant state) of the subsonic region multiplied by a common phase factor $\exp\left[\rmi S(x,t)|\right]$ that fixes the initial hydrodynamic velocity field.
In the limit of a negligible barrier size the time-dependent potential is equivalent to the boundary conditions $\psi(x_{{\rm piston}},t)=\psi(0,t)=0$. Therefore the $\psi_{\rm n} (x,t)$ can be approximated in the initial stage of the dynamics by 
\begin{equation}
\psi_{\rm n} (x,t) = \frac1{\sqrt{2\,|x_{\rm piston}(t)|}}\sin\left[\frac{n\,\pi\,x}{|x_{\rm piston}(t)|}\right] \; \exp\left[-\rmi \, \frac{m\,v_{\rm piston}\, x^2}{2 \hbar\,|x_{\rm piston}(t)|}\right]
\label{psin}
\end{equation}
which are exact solutions of the Schr{\"o}dinger equation with the above specified boundary conditions.
Note that as the piston evolves towards the barrier the number of nodes $n$ of $\psi_{\rm n}$ is conserved in time and its effective wavelength grows.
The semiclassical velocity distribution of (\ref{flow2}) with the wave-functions $\psi_{\rm n} (x,t)$ given by (\ref{psin}) corresponds
indeed to the initial distribution given previously by equation (\ref{fxvt}) and equations (\ref{vli}), (\ref{vri}), (\ref{pi}) and (\ref{vfi}) as one can easily check.

As the piston moves even closer to the barrier the wave-functions
$\psi_{\rm n} (x,t)$ start to propagate one by one over the barrier. The differences in energy levels in the subsonic region
are assumed to be much less that the Hawking energy, i.e. $h\,v_{\rm esc} / 2 |x_{\rm piston}| \ll k_{{\rm B}}\,T_{{\rm H}} $ where $T_{{\rm H}}$ is given by equation (\ref{bare1}). We can thus use the transmission formula (\ref{penetration}) to predict the distribution of momenta in the supersonic region. The predictions will be compared with the results obtained by the numerical integration of the Schr{\"o}dinger equation in subsection 3.3.

Consider a particle transmitted in the supersonic region with some velocity $v_{\rm final}$. Before exit the subsonic region it has been reflected between the piston and the barrier a certain number of times. Therefore the probability $p$ of this process to occur is proportional to
\begin{eqnarray}
p(v_{\rm final}) &\propto& |t(v_{\rm final})|^{2}\,|r(v_{\rm final}-2\,v_{\rm piston})|^{2}\,...\,|r(v_{\rm final}-2\,n_{\rm coll}\,v_{\rm piston})|^{2}\nonumber\\
&\propto& |t(v_{\rm final})|^{2}\,\Pi_{n=1}^{n_{\rm coll}} |r(v_{\rm final}-2\,n\,v_{\rm piston})|^{2}\label{probability}
\end{eqnarray}
where $n_{\rm coll}$ is the number of collisions with the barrier and $|t|^{2}$ and  $|r|^{2}$ are the transmission and reflection probabilities as function of the velocity.
In the stationary regime, the final distribution of velocity $f(v)$ should not depend anymore on the initial distribution and it should be given, a part from a normalization constant, by
\begin{eqnarray}
f(v) = |t(v)|^{2}\,\Pi_{n=1}^{n_{\rm upper}} |r(v-2\,n\,v_{\rm piston})|^{2}\label{distribution}
\end{eqnarray}
where $n_{\rm upper}$ is an integer for which  $|r(v_{\rm esc}-2\,n_{\rm upper}\,v_{\rm piston})|^{2}=1$.

In particular, when the Hawking energy is much smaller than the difference between left and right Fermi energies in the stationary regime, i.e. $k_{{\rm B}}\,T_{{\rm H}} \ll m\,v_{\rm piston}\,v_{\rm esc} $ equation (\ref{distribution}) simplifies to
\begin{eqnarray}
f(v) = |t(v)|^{2}\,|r(v-2\,v_{\rm piston})|^{2}
\label{distributionsimple}
\end{eqnarray}
as the next terms of the product in (\ref{distribution}) are all equal to one. 
This condition is equivalent to the condition  $k_{{\rm B}}\,T_{{\rm H}} \ll m c^{2}$ where $c$ is the speed of sound on the event horizon.   
Thus in the supersonic region, the occupation probability $p(\epsilon)$ of an energy state $\epsilon$ is given a part from a normalization factor, by
\begin{eqnarray}
p(\epsilon)&=& \frac{\exp((\epsilon-V_{\rm max} )/k_{\rm B}T^{\rm 0}_{\rm H})}{\left[1+\exp((\epsilon-V_{\rm max} )/k_{\rm B}T^{\rm 0}_{\rm H})\right]\left[1+\exp((\epsilon'-V_{\rm max} )/k_{\rm B}T^{\rm 0}_{\rm H})\right]}
\label{probabilityeps}
\end{eqnarray}
where $\epsilon'$ is the energy of the particle before colliding to the piston
\begin{equation}
\epsilon'(\epsilon)=m(\sqrt{2\epsilon/m}-2 v_{\rm piston})^{2}/2
\label{epsilon'}
\end{equation}
Therefore differently from the case discussed in Section 2 both the left and right Fermi points $v_{\rm L}(x)$ and $v_{\rm R}(x)$ are smoothened by the tunnelling through the barrier. The reason of this is that $v_{\rm L}$ and $v_{\rm R}$ are recombined by the piston (see Figure 3). Remarkable the gas appears as thermally distributed on both sides of the local Fermi velocity distribution.

\subsection{Numerical results}

We would like to compare the above theory with a numerical integration of the Schr{\"o}dinger equation for the many-particle wave-function of the gas (\ref{flow2}). However, even if the system is non-interacting and therefore treatable in principle we should note that the above conclusions are valid for a very large system. 
Different requirements need to be simultaneously satisfied: (i) smooth  and relatively extended barrier; (ii) $|x_{\rm piston}|$ much larger than the barrier extension to minimize finite size and transient effects; (iii) large number of fermions; (iv) long time scale as the piston moves slowly towards the barrier.

On the other hand, the time evolutions of the wave-functions $\psi_{\rm n}$ are quite similar.
Each $\psi_{\rm n}$ defined in (\ref{psin}) is a  superposition of $\exp( \rmi \,n\,x\,/\,|x_{\rm piston}| -\rmi \, m\,v_{\rm piston}\, x^2 / 2 \hbar \,|x_{\rm piston}| )$ and $\exp(- \rmi \,n\,x\,/\,|x_{\rm piston}| -\rmi \, m\,v_{\rm piston}\, x^2 / 2 \hbar \,|x_{\rm piston}| )$, which contain classical velocities uniformly distributed in the intervals $(h\,n\,/\,m \,|x_{\rm piston}|,h\,n\,/\,m \,|x_{\rm piston}|+v_{\rm piston})$ and $(-h\,n\,/\,m \,|x_{\rm piston}|,-h\,n\,/\,m \,|x_{\rm piston}|+v_{\rm piston})$, respectively, where $h$ is the Planck constant. Each $\psi_{\rm n}$ performs thus a reasonable sampling of the phase space distribution of the gas.
We therefore study the time-evolution of only one of the $\psi_{\rm n}$.

We calculate the stationary momentum distribution of the flow (\ref{flow2}), which is a Slater determinant by approximating the sum of single-particle contributions by only the contribution of one  of the $\psi_{\rm n}$.
Its momentum distribution in the supersonic region is calculated 
from the frequency components of $\psi_{\rm n}(r_{\rm ref},t)$ at some arbitrary reference point $r_{\rm ref}$ in the supersonic region where the particles propagate as a plane waves and taking into account the density of states.

Figure 4 shows some examples of the comparison between the momentum distribution in the supersonic region calculated from one of the $\psi_{\rm n}$ and equation (\ref{distribution}) with (\ref{bare1}) as formula for the reflection probability.  The agreement is reasonable considering that the chosen initial wave functions have only $n=14$ nodes. $\psi_{\rm n}$ is a quasi-resonant state of the left region  multiplied by the phase factor $\exp\left[-\rmi \, m\,v_{\rm piston}\, x^2 / 2 \hbar\,|x_{\rm piston}(0)|\right]$, which gives a constant gradient of velocity field decreasing from the maximum value $|v_{\rm piston}|$ at the classical turning point of $\psi_{\rm n}$ at $x_{\rm piston}$ to zero near the barrier.

\begin{figure}[htbp]
\includegraphics[width=80mm]{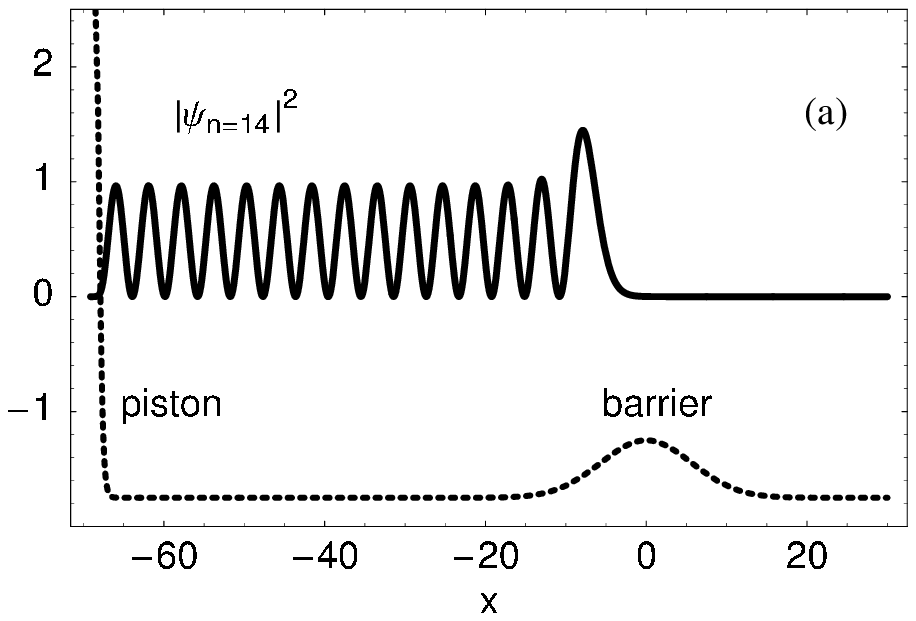}
\includegraphics[width=80mm]{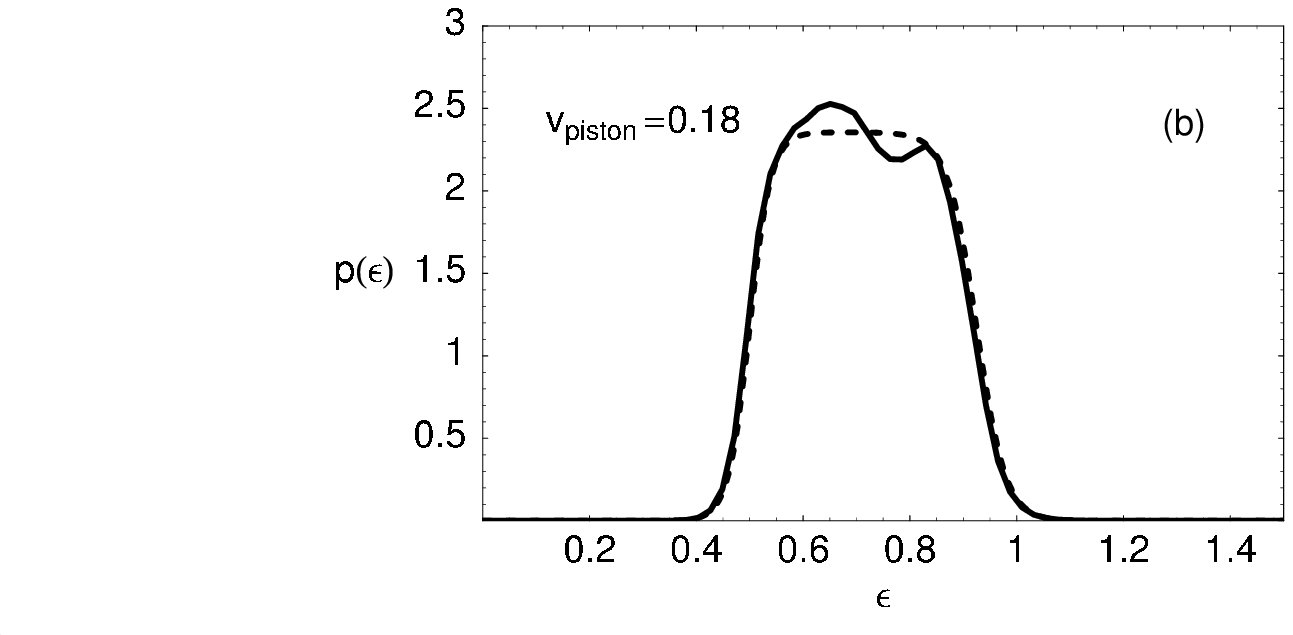}
\includegraphics[width=80mm]{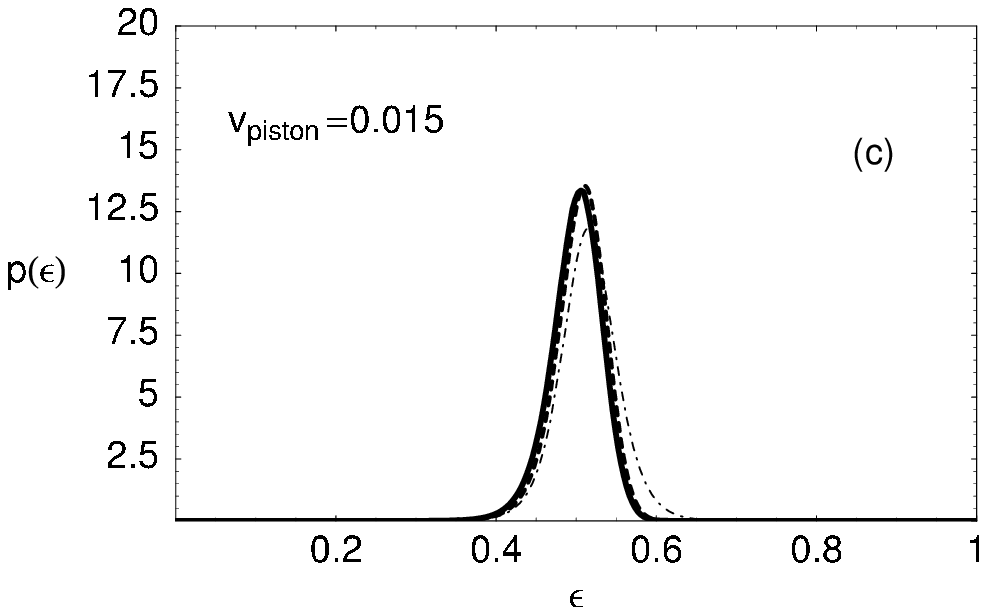}
\includegraphics[width=80mm]{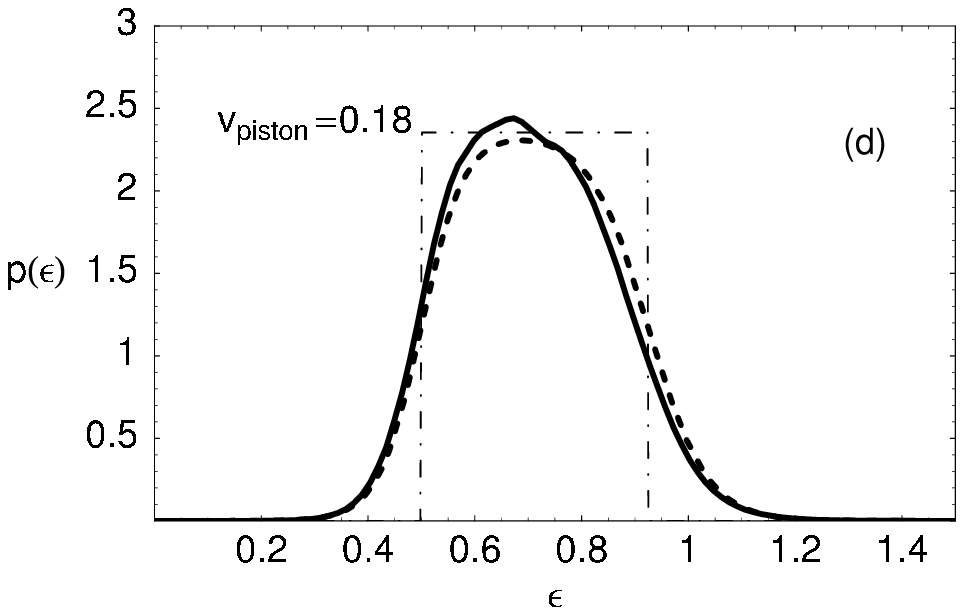}
\caption{(a) Example of numerically computed density of an initial resonant state with $n=14$ nodes (solid line) and potential energy (dashed line) at the initial time of the evolution. In dimensionless units $V_{\rm max}=1/2$, $v_{\rm esc}=1$ and $k_{B} T_{H}^{0}=0.02$. Figures (b), (c), (d)  show  the occupation probabilities  in the supersonic region of plane waves with energy $\epsilon$ obtained by numerical integration of the Schr{\"o}dinger equation (solid lines) and from the analytical formulas (\ref{distribution}) and (\ref{bare1}) (dashed lines). 
In figure (b) and (d) ($k_{B} T_{H}^{0}=0.02$ and $k_{B} T_{H}^{0}=0.04$, respectively) $v_{\rm piston}=0.18$ and the dashed line is calculated using (\ref{distributionsimple}) where the multiple reflections corrections (\ref{distribution}) are neglected. 
In figure (c) $v_{\rm piston}=0.015$ is chosen such that the Hawking energy $k_{B} T_{H}^{0}=0.02$  is comparable to $m c^{2}$ at the horizon and the multiple reflections corrections (\ref{distribution}) (dashed line) are necessary. Dot-dashed line is obtained from (\ref{distributionsimple}). 
Note that in all three examples (b), (c), (d) the classical gas distribution is always a sharply defined interval showed for comparison with the dot-dashed line only in (c).
The small deviations between solid and dashed lines 
are quite sensitive on small changes in the exponential phase factor (not shown here) and on the number of nodes of the initial wave function, which is $n=14$ in all three examples.}
\label{figure1}\end{figure}

\section{Hawking radiation in the laboratory: Relevant experimental parameters} 
Today's atomic trapping techniques are sufficiently developed to ensure the realization of a setup for the demonstration in the laboratory of the sonic analogue of Hawking radiation. The main difficulty is the realization of an ``event horizon'' barrier that allows appreciable tunnelling. Josephson-like effects in Bose-Einstein condensates \cite{Giovanazzi97,Giovanazzi00a} have been recently observed \cite{Oberthaler05}. This proves that it is possible to achieve barriers that allow a reasonable tunnelling.
Here follows an estimation for a Fermi degenerate gas of Lithium atoms, which has been recently obtained \cite{Lithium}. Lithium can have a degeneracy temperature as high as $8\;\mu$K, which fixes the scale of $\epsilon_{\rm max}$. The barrier potential could be created by a far-blue-detuned laser sheet propagating in a direction orthogonal to that of the flow propagation. One of the resonance lines of Lithium is at $233$ nm \cite{nist}. Using an ultra-violet diode laser, which is commercially available and tightly focused with a waist ($1/e^{2}$) of twice the laser wavelength (that can be tuned at $233$ nm) the Hawking temperature is of order of $T_{{\rm H}} \sim 2 $ $\mu$K, which is a reasonable value and comparable with the degeneracy temperature of $8\;\mu$K. The characteristic wavelength of Hawking radiation $\lambda_{c}$ is of order of $4\;\mu$m.

\section{Conclusions}

The exactly solvable model, originally formulated in terms of one-dimensional scattering of a Fermi gas \cite{Giovanazzi05a} is extended to account for a finite size realistic geometry. The flow of particles is maintained by a piston moving slowly towards the sonic horizon. 
Both points of the Fermi segment appear thermally smoothed.
Moreover multiple reflections between piston and barrier lead to possible modifications of the result obtained in \cite{Giovanazzi05a}
for the momentum distribution.
Interestingly, due to the geometry of the moving piston the prediction on the thermal momentum distribution can be tested studying the time-evolution of only one single-particle wave-function.

We estimated that by using existing technologies the Hawking temperature can be of order of a few microkelvin in a realistic experiment, which is a reasonable temperature to be measured.

\ack
I thank U. Leonhardt and T. Kiss for inspiring discussions on sonic black holes in atomic Bose-Einstein condensates, W. Unruh for a discussion on the hydrodynamic of the one-dimensional degenerate gas, D. Cassettari, J. Fortagh, S. Hensler, T. Pfau and C. Zimmermann for discussions especially on the experimental feasibility of the proposed experiment. I acknowledge the Marie Curie Programme of the European Commission and the German Science Foundation (DFG) (SFB/TR 21) for support.

\end{document}